\documentclass{iopconfser}

\usepackage[utf8]{inputenc}
\usepackage{graphicx}
\usepackage{amsmath,amssymb,amsfonts}
\usepackage{caption}
\usepackage{subcaption}
\usepackage{ragged2e}

\begin{document}

\title{Soliton Dynamics of a Gauged Fokas-Lenells Equation Under Varying Effects of Dispersion and Nonlinearity}

\author{Riki Dutta$^{1*}$, Sagardeep Talukdar$^{1}$, Gautam K. Saharia$^{1}$, and Sudipta Nandy$^{1}$}

\affil{$^1$Department of Physics, Cotton University, Guwahati, India}

\email{rikidutta96@gmail.com}

\begin{abstract}
\justifying
Davydova-Lashkin-Fokas-Lenells equation (DLFLE) is a gauged equivalent form of Fokas-Lenells equation (FLE) that addresses both spatio-temporal dispersion (STD) and nonlinear dispersion (ND) effects. The balance between those effects results a soliton which has always been an interesting topic in research due to its potential applicability as signal carrier in information technology. We have induced a variation to the dispersion effects and apply Hirota bilinear method to realise soliton solution of the proposed DLFLE and explore how the soliton dynamic behaves in accordance to the variation of the dispersion effects. The proposed equation is applicable for number of systems like ultrashort optical pulse, ion-cyclotron plasma wave, Bose-Einstein condensate (BEC) matter-wave soliton under certain external fields, etc. The study on such systems under varying effects is very limited and we hope our work can benefit the researchers to understand soliton dynamics more and work on various other nonlinear fields under varying effects.
\end{abstract}
\vspace{10mm}

\section{Introduction}
\justifying
Soliton is a nonlinear wave structure that arises as a result of balance between dispersion and nonlinear effects of the medium. Soliton can travel long distances without getting distorted and therefore making it a suitable signal carrier in information technology. Since the discovery of soliton, the research on this topic is highly active. Nonlinear Schrödinger equation (NLSE) is the most basic example of nonlinear equations that considers both group velocity dispersion (GVD) and Kerr nonlinear effect (KNE), the balance between those is required for the existence of soliton. NLSE is applicable in most of the Kerr nonlinear media namely, nonlinear optics, BEC, plasma, etc. with appropriate modifications. The solution of NLSE in nonlinear optics is an optical soliton \cite{malomed1996soliton, hosseini2023dynamics}. In case of BEC, NLSE is popularly known as Gross-Pitaevskii equation (GPE) which contains kinetic energy operator and inter bosonic interactions as analogous to GVD and KNE respectively. The solution of GPE is a matter-wave soliton \cite{atre2006class, al2009integrability}. In various fields of physics, soliton remains an interesting topic as it also tells about the integrability of the system. Beyond GVD and KNE, asymptotic terms like STD and ND also exist but are neglected since in most of the cases they do not alter the dynamics of the soliton. However, in some cases, like for an ultrashort pulse, ion-cyclotron plasma wave, BEC under some certain external potentials, we need to address these effects as the existence of these two terms effect the dynamics of the soliton. FLE is a partial differential equation that contains all these effects and the dimensionless integrable form of FLE \cite{lenells2008novel, lenells2009exactly} can be expressed as
\begin{align} 
	\label{FLE}
	iU_t  + a_1 U_{xx} - a_2 U_{xt} + b |U|^2\ ( U + i\ a_2 U_x) &= 0 
\end{align}
where $U$ is the field function that describes the complex waveform of the optical pulse or matter-wave structure, etc. The suffix $x$ and $t$ denote the partial differentiation of $U$ with respect to $x$ and $t$ respectively. $U_t$ is the temporal evolution of the waveform and $U_{xx}$ and $|U|^2 U$ represent GVD and KNE respectively. In nonlinear optics and plasma physics, $U_{xt}$ and $|U|^2 U_x$ represent STD and ND respectively. In BEC however, the introduction of ND term into GPE is by the phase-imprinting technique (PIT) under some properly designed optical field \cite{kengne2013phase, ramesh2010phase}. In the quantum context, STD ($U_{xt}$) represents a temporal evaluation of the momentum of the waveform or in other words the negative flow of potential, i.e. force. For $a_2 = 0$, Eq. (\ref{FLE}) is nothing but NLSE (or GPE). In addition, we introduce a varying dispersion and nonlinear effects. Our initial system represented by Eq. \ref{FLE} modifies into the following form
\begin{align}
	\label{mFLE}
	iU_t  + a(t) a_1\ U_{xx} - a_2 U_{xt} + a(t) b |U|^2\ (U + i\ a_2 U_x) &= 0
\end{align}
here the time-varying coefficient $a(t)$ represents variation in GVD and nonlinear effects. The study on FLE system is relatively new \cite{ebaid2019exact, hendi2021dynamical, biswas2018optical, dutta2023fokas, talukdar2023multi, talukdar2023linear, saharia2024data} and very limited works have been published under varying dispersion and nonlinear effect \cite{kundu2010two, lu2013nonautonomous, dutta2024soliton}. The study on such a system holds very good scope to understand the insights of the systems and provides a control over the soliton management to tune the characteristics of the soliton according to our desire. Our focus in this paper is oriented towards the understanding of the soliton dynamics due to the presence of STD and ND under varying effects.
\vspace{2mm}

At first, we transform Eq. \ref{mFLE} into a simplified form called DLFLE which contains STD and ND along with the varying coefficient. Then we apply Hirota bilinearisation to DLFLE to solve for one-soliton solution (1SS) and two-soliton solution (2SS) and graphically represent the obtained soliton waveforms. We also explore the soliton dynamics for various forms of the varying coefficients. The structure of this paper is as follows: in the following section we convert FLE (Eq. \ref{mFLE}) under a gauge transformation followed by a variable transformation to get the DLFLE form (Eq. \ref{DLFLE}). In the same section we apply Hirota bilinear method and transform our higher order nonlinear equation into three bilinear forms. In the third section we proceed to obtain the soliton solutions of DLFLE. After obtaining the soliton solutions, we present the solitons graphically under various forms of the varying coefficient and observe the dynamics of solitons accordingly. In the fourth section we conclude our paper.

\section{Hirota Bilinear Method}
\justifying
Since the constant coefficients ($a_1$, $a_2$, $b$) can be absorbed under some particular gauge transformation so the choice for their values can be considered as per our convenience. Assuming $n=\frac{1}{a_2} $, $m=\frac{a_1}{a_2} > 0$, consider the gauge transformation 
\begin{align}
	\label{GT}
	U = \sqrt{\frac{m}{|b|}}n e^{i(n x +2 mn \int{a(t)} dt)} u 
\end{align}
followed by the transformation of variables
\begin{align}
	\label{VT}
	\xi = x+m \int{a(t)} dt, \quad \tau = -mn^2 t
\end{align}
and for the time-varying coefficient consider
\begin{align}
	a(t) \rightarrow a(\tau)
\end{align}
then the Eq. \ref{mFLE} transforms into the following form
\begin{align} 
	\label{DLFLE1}
	u_{\xi \tau} - a(\tau)\ u + i\sigma\ a(\tau)\ |u|^2 u_\xi &= 0
\end{align}
here $u$ is the field function corresponding to the new transformed system and $\sigma = sign(b) = \pm 1$. Under a simple variable transformation: $\xi \to \sigma \xi$ and $\tau \to \frac{\tau}{\sigma}$, we can omit $\sigma$ from Eq. \ref{DLFLE1} and can be rewritten as
\begin{align} 
	\label{DLFLE}
	u_{\xi \tau} - a(\tau)\ u + i\ a(\tau)\ |u|^2 u_\xi &= 0
\end{align}

For constant dispersion and nonlinear effects, i.e. constant coefficient ($a(\tau)=1$), Eq. (\ref{DLFLE}) becomes the first negative hierarchy of derivative NLSE and is similar to an equation given by Davydova and Lashkin which governs the dynamics of short-wavelength ion-cyclotron waves in plasma \cite{davydova1991short, lashkin2021perturbation}. In our paper we refer Eq. \ref{DLFLE} as DLFLE and our study orients towards this equation.

\vspace{2mm}
Now under a vanishing background condition, i.e. $u \to 0$ as $\xi \to \pm \infty$ we transform DLFLE into three bilinear forms. For this we assume $u$ to take the following form
\begin{align}
	\label{ugf}
	u &= \frac{g}{f}
\end{align}
where $g$ and $f$ are functions of $\xi$ and $\tau$. Putting Eq. \ref{ugf} into Eq. \ref{DLFLE} we get the following expression
\begin{align}
	\label{bilin}
	\frac{1}{f^2}(D_\xi D_\tau - a(\tau)) g.f - \frac{g}{f^3}D_\xi D_\tau(f.f) +  \frac{2i\ |g|^2}{f^3 f^*} a(\tau) D_\xi (g.f) +
	\frac{s |g|^2}{f^3} - \frac{s |g|^2 f^*}{f^3 f^*} =0
\end{align}
where $D_\xi$, $D_\tau $ are Hirota derivatives \cite{hietarinta2007introduction, hirota1976n} and are defined as
\begin{align}
	D_\xi^m D_\tau^n g(\xi,\tau).f(\xi,\tau)= 
	(\frac{\partial}{\partial \xi}  - \frac{\partial}{\partial \xi^\prime})^m
	(\frac{\partial}{\partial \tau}  - \frac{\partial}{\partial \tau^\prime})^n
	g(\xi,\tau).f(\xi^\prime,\tau^\prime)\Bigg|_{ (\xi=\xi^\prime)(\tau= \tau^\prime)}
\end{align} 

The last two terms in Eq. \ref{bilin} contains an auxiliary function ($s$) which is introduced to get three bilinear equations as below
\begin{align}
	\label{BR1}
	(D_{\xi} D_{\tau} - a(\tau))g.f &= 0 \\
	\label{BR2}
	D_{\xi} D_{\tau}f.f  &= sg^* \\
	\label{BR3}
	i\ a(\tau) D_{\xi} (g.f) &= sf^* 
\end{align}

To obtain the soliton solution,  $g$ and $f$ are expanded with respect to an arbitrary parameter $\epsilon$ as follow
\begin{align}
	\label{gf}
	g&= \epsilon g_1 + \epsilon^3 g_3 + ..., \quad \quad \quad 
	f = f_0 + \epsilon^2 f_2 + \epsilon^4 f_4 + ...
\end{align}
and the auxiliary function $s$ is expanded as  
\begin{align}
	\label{s}
	s = \epsilon s_1 + \epsilon^3 s_3 + ...
\end{align}  

In the following section we proceed to realise the 1SS and 2SS of DLFLE.

\section{Solution of DLFLE}
\subsection{1SS of DLFLE}

\begin{figure}
	\centering
	\begin{subfigure}{.3\textwidth}
		\includegraphics[width=\textwidth]{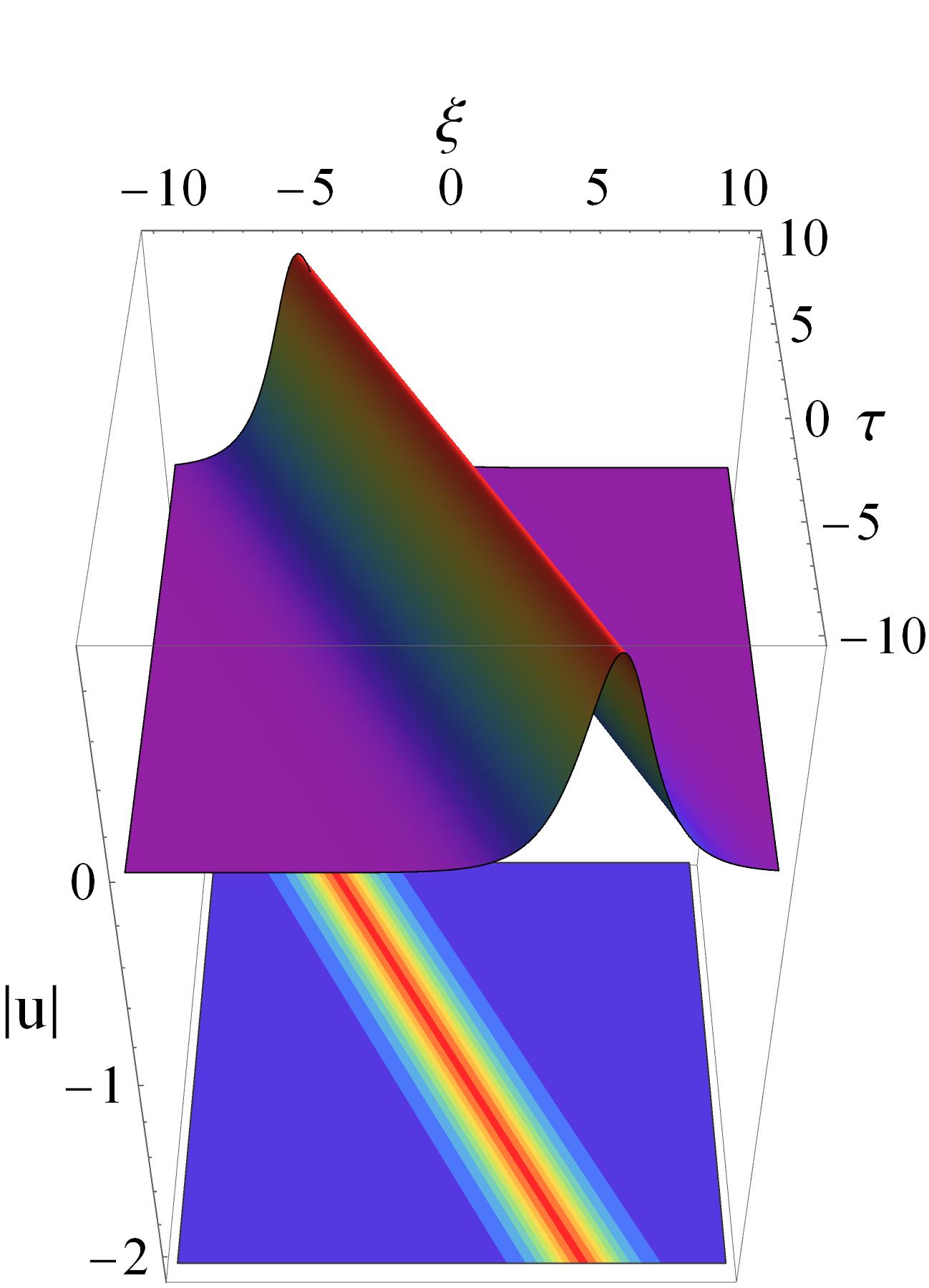}
		\caption{$a(\tau) = 1$}
		\label{Fig1a}
	\end{subfigure}
	\begin{subfigure}{.3\textwidth}
		\includegraphics[width=\textwidth]{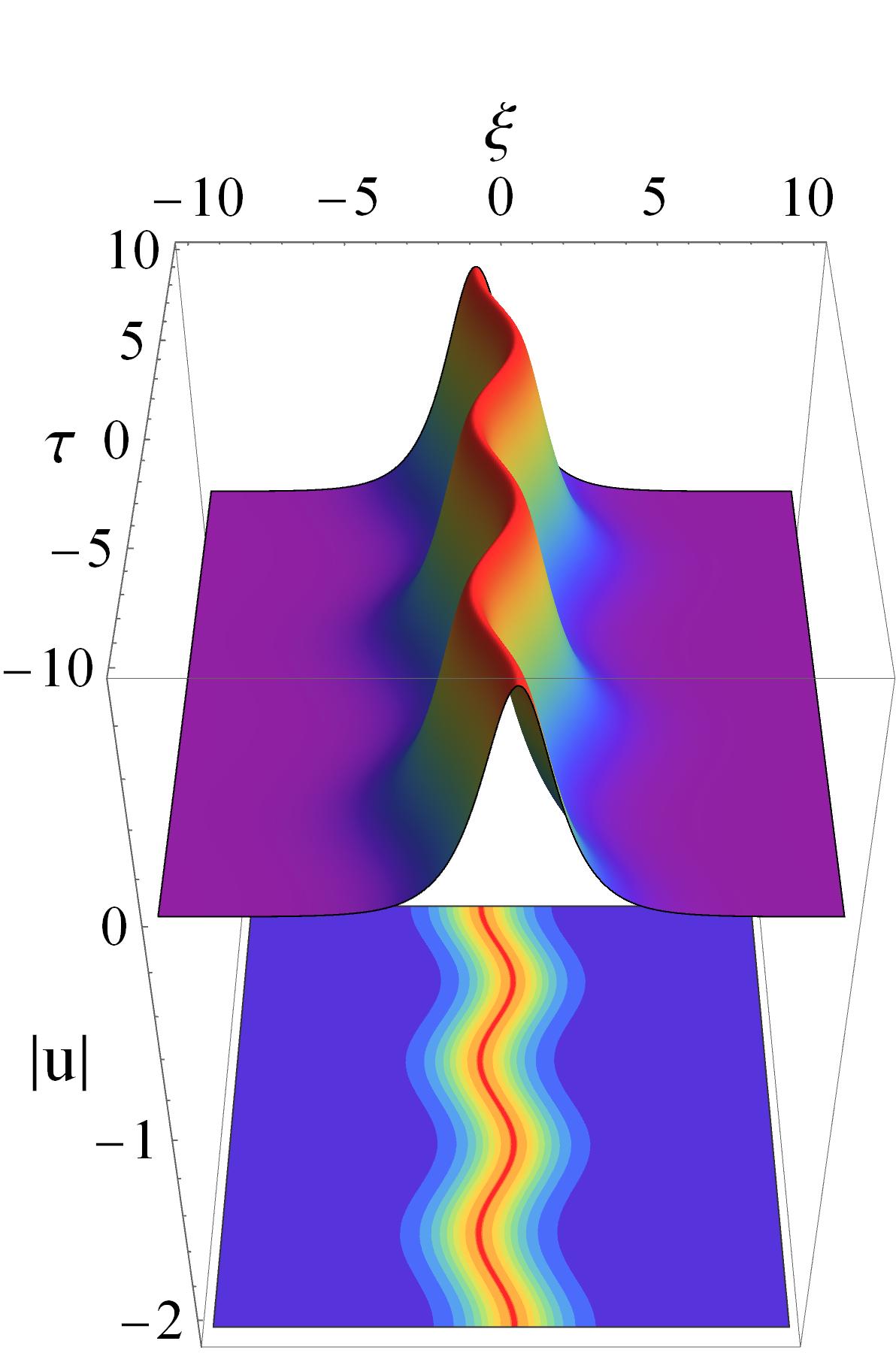}
		\caption{$a(\tau) = cos(k t)$}
		\label{Fig1b}
	\end{subfigure}
	\begin{subfigure}{.3\textwidth}
		\includegraphics[width=\textwidth]{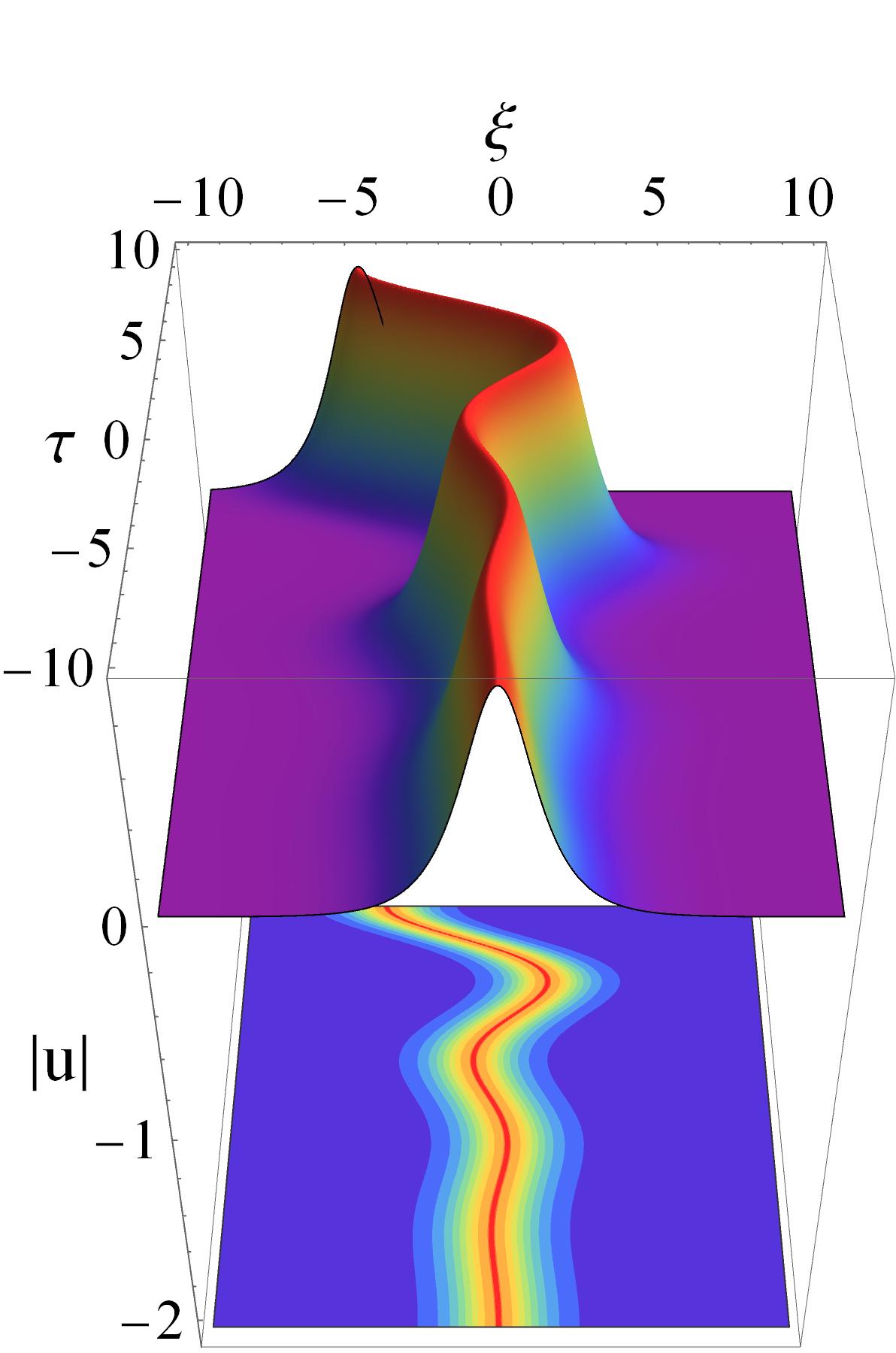}
		\caption{$a(\tau) = e^{\sigma t} cos(k t)$}
		\label{Fig1c}
	\end{subfigure}
	\caption{All the three graphs represent 1SS of DLFLE under three different forms of the varying coefficient. (a) $a(\tau) = 1$, (b) $a(\tau) = cos(k \tau)$ and (c) $a(\tau) = e^{\sigma \tau} cos(k \tau)$. We have fixed $c_1 = \sqrt{-1}$, $p_1 = 1 + i$, $k = \frac{\pi}{4}$ and $\sigma = 0.2$}
	\label{Fig1}
\end{figure}

For 1SS of DLFLE we ignore the terms of order $\epsilon^3$ and higher from the expressions of $g$, $f$ and $s$ in Eqs. \ref{gf} and \ref{s}. Now the expressions for  $g$, $f$ and $s$ can be expressed as follow
\begin{align}
	\label{gfs1}
	g = g_1, \quad \quad f = f_0 + f_2,  \quad \quad s = s_1
\end{align}
since $\epsilon$ is introduced to represent the order of the terms so we omit it from the above expressions. This makes the expression of Eq. \ref{ugf} for 1SS as
\begin{align}
	\label{ugf1}
	u &= \frac{g_1}{1 + f_2}
\end{align}
Eq. \ref{ugf1} is the 1SS of DLFLE (Eq. \ref{DLFLE}). Now considering the following expressions
\begin{align}
	\label{g1}
	g = \alpha_1 e^{\Theta_1} \\
	\label{f1}
	f = 1 + \beta_2 e^{\Theta_1 + \Theta_1^*} \\
	\label{s1}
	s = \gamma_1 e^{\Theta_1}
\end{align}
where $\Theta_1 = p_1 \xi + \frac{1}{p_1} \int a(\tau) d\tau$. $p_1$ is the spectral parameter. Putting Eqs. \ref{g1} - \ref{s1} into Eqs. \ref{BR1} - \ref{BR3}, we get the expressions of $\alpha_1$ and $\beta_2$ as
\begin{align}
	\label{beta2}
	\beta_2 &= i c_1^2 p_1 \\
	\label{alpha1}
	\alpha_1 &= \sqrt{2} c_1 (1 + \frac{p_1^*}{p_1})
\end{align}
where $c_1$ is an arbitrary real or purely imaginary constant. Now we have all the expressions needed to represent 1SS graphically. Figures \ref{Fig1a} - \ref{Fig1c} represent the same. Three graphs are presented for three different cases of the varying coefficient $a(\tau)$. 1. Figure \ref{Fig1a}: $a(\tau) = 1$ represents constant dispersion and nonlinear effects i.e. no variation which results a straight propagation of the soliton, 2. Figure \ref{Fig1b}: $a(\tau) = cos(k \tau)$ represents a sinusoidal variation and the propagation of the soliton is oscillating about the mean position $\xi = 0$ and 3. Figure \ref{Fig1c} $a(\tau) = e^{\sigma \tau} cos(k \tau)$ represents an exponentially increasing sinusoidal function which results increasing deviation of the soliton oscillation from mean position.
\vspace{2mm}

The width of the 1SS is expressed as $\frac{2}{p_1 + p_1^*}$ and the velocity is $\frac{a(\tau)}{|p_1|^2}$. The expression of velocity explains why we observe sinusoidal propagation of the soliton in Figures \ref{Fig1b} and \ref{Fig1c} as the velocity is itself a function of the varying coefficient $a(\tau)$. This concludes that the choice of $a(\tau)$ can directly alter the dynamics of the soliton and giving us a good control over the soliton management.
\vspace{2mm}

If readers are interested then they may realise the 1SS of FLE (Eqs. \ref{FLE} and \ref{mFLE}) by putting Eq. \ref{ugf1} into Eq. \ref{GT} and considering Eq. \ref{VT}.

\subsection{2SS of DLFLE}

\begin{figure}
	\centering
	\begin{subfigure}{.3\textwidth}
		\includegraphics[width=\textwidth]{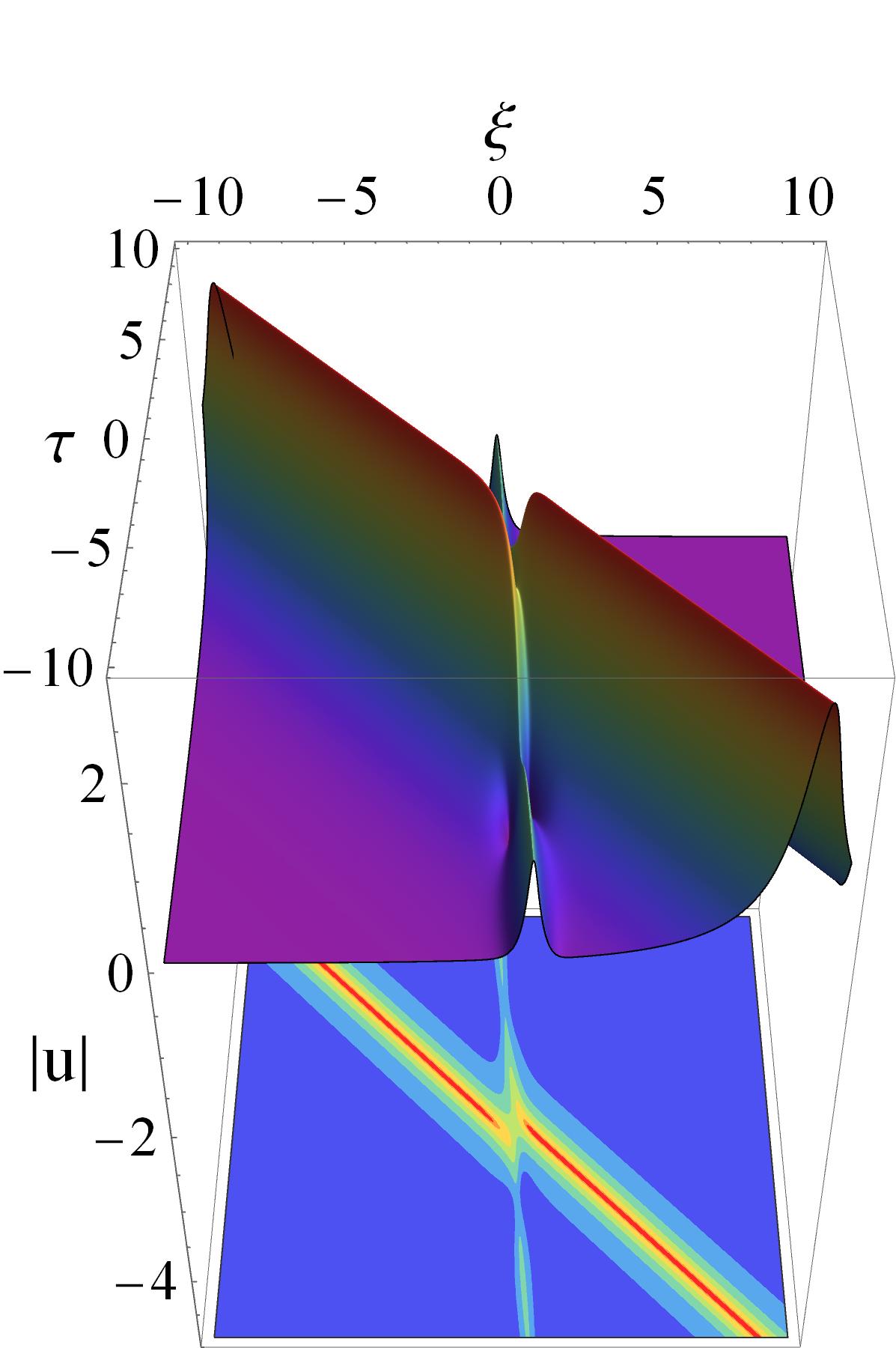}
		\caption{$a(\tau) = 1$}
		\label{Fig2a}
	\end{subfigure}
	\begin{subfigure}{.3\textwidth}
		\includegraphics[width=\textwidth]{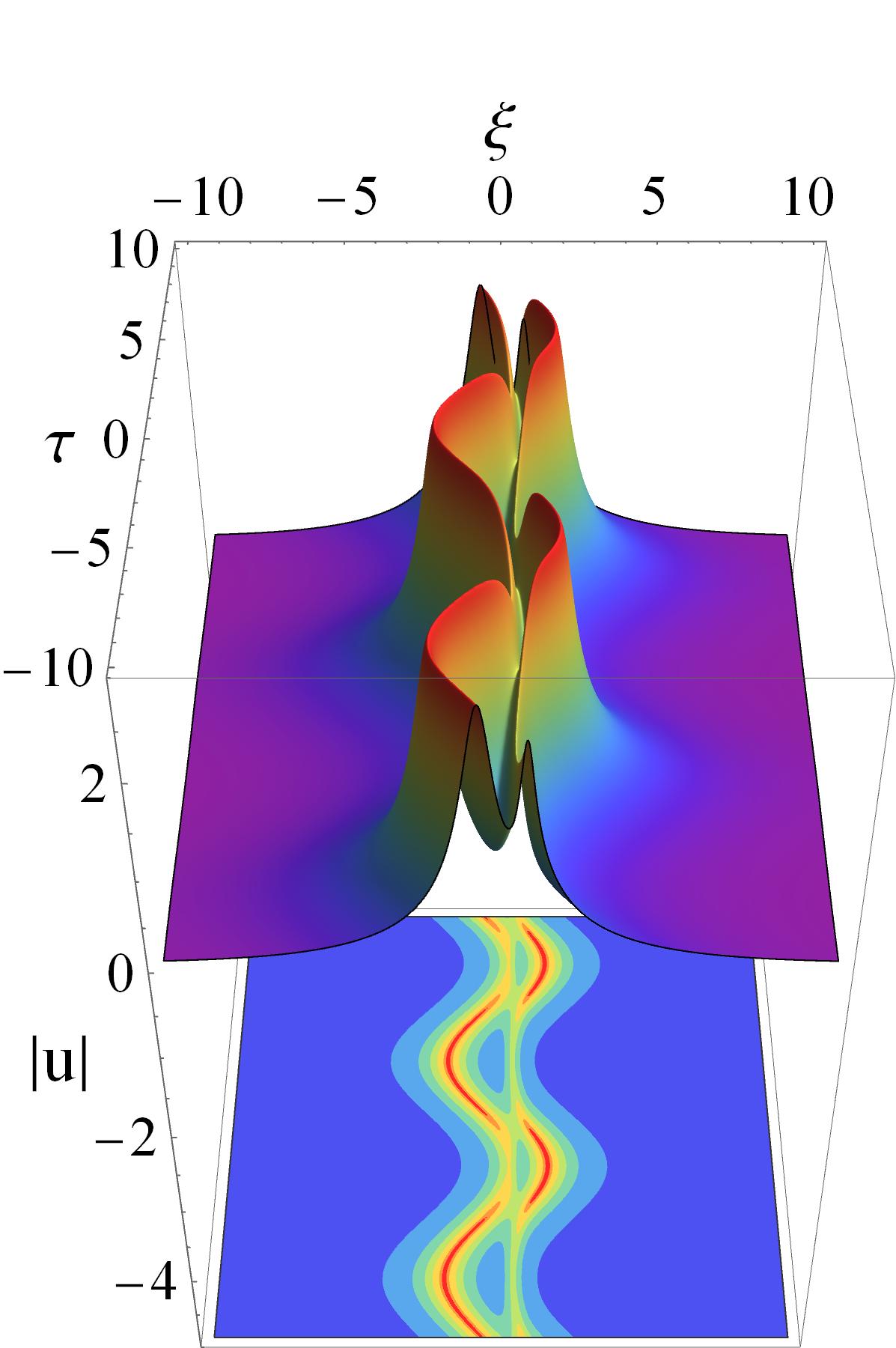}
		\caption{$a(\tau) = cos(k t)$}
		\label{Fig2b}
	\end{subfigure}
	\begin{subfigure}{.3\textwidth}
		\includegraphics[width=\textwidth]{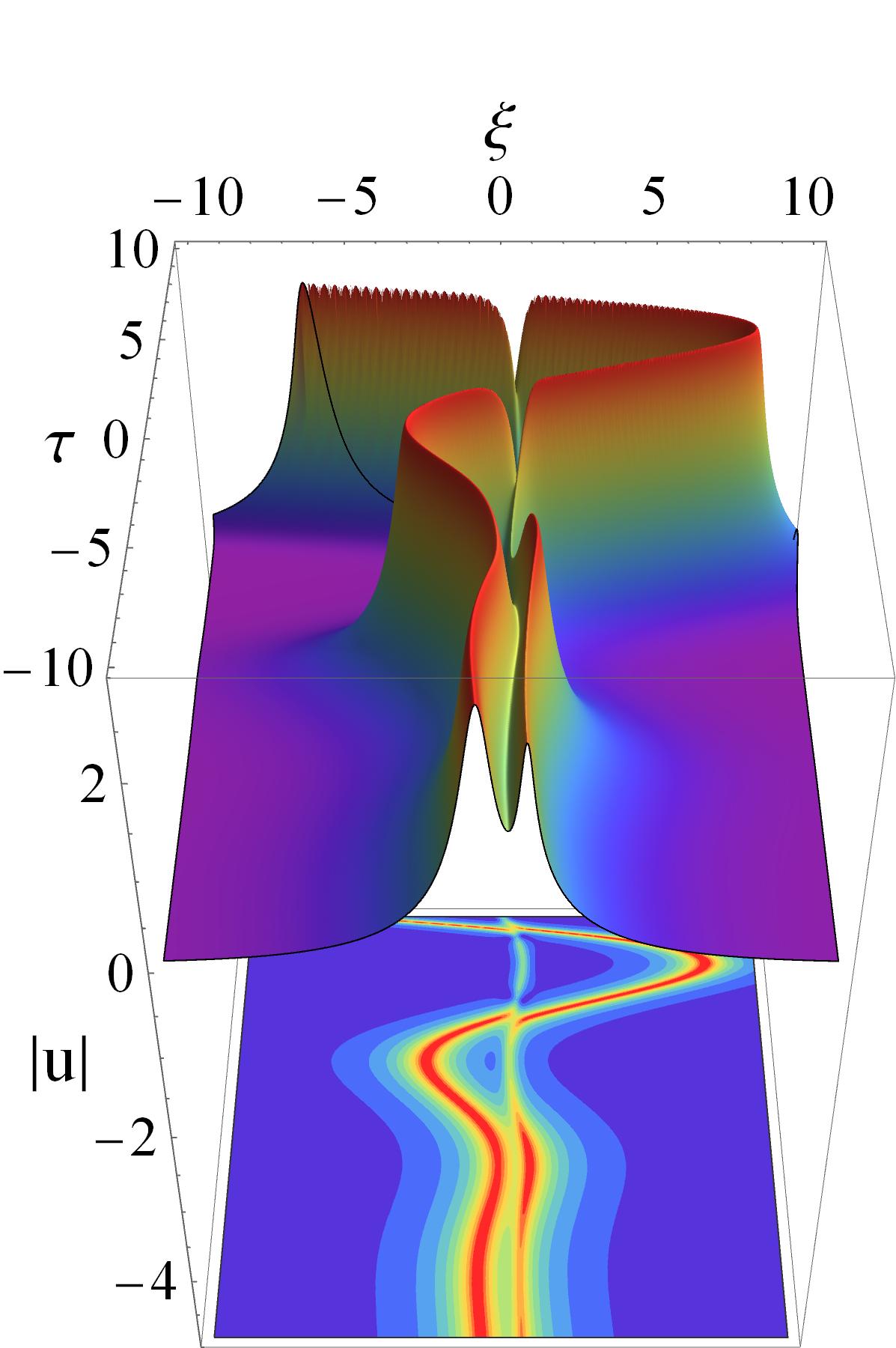}
		\caption{$a(\tau) = e^{\sigma t} cos(k t)$}
		\label{Fig2c}
	\end{subfigure}
	\caption{All the three graphs represent 2SS of DLFLE under three different forms of the varying coefficient. (a) $a(\tau) = 1$, (b) $a(\tau) = cos(k \tau)$ and (c) $a(\tau) = e^{\sigma \tau} cos(k \tau)$. We have fixed $c_2 = 1$, $p_1 = 0.4 + i$, $p_2 = 4 + i$, $k = \frac{\pi}{5}$ and $\sigma = 0.25$.}
	\label{Fig2}
\end{figure}

For 2SS of DLFLE we ignore the terms of order $\epsilon^5$ and higher from the expressions of $g$, $f$ and $s$ in Eqs. \ref{gf} and \ref{s}. Now the expressions for  $g$, $f$ and $s$ can be expressed as follow
\begin{align}
	\label{gfs2}
	g = g_1 + g_3, \quad \quad f = 1 + f_2 + f_4,  \quad \quad s = s_1 + s_3
\end{align}
This makes the expression of Eq. \ref{ugf} for 2SS as
\begin{align}
	\label{ugf2}
	u &= \frac{g_1 + g_3}{1 + f_2 + f_4}
\end{align}
Eq. \ref{ugf2} is the 2SS of DLFLE. Now considering the following forms
\begin{align}
	\label{g21}
	g_1 = \alpha_1 e^{\Theta_1} + \alpha_2 e^{\Theta_2}\\
	\label{g23}
	g_3 = \alpha_{12} e^{\Theta_1 + \Theta_1^* + \Theta_2} + \alpha_{21} e^{\Theta_1 + \Theta_2 + \Theta_2^*}\\
	\label{f22}
	f_2 = \sum_{j, k = 1}^{2} \beta_{jk} e^{\Theta_j + \Theta_k^*} \\
	\label{f24}
	f_4 = \beta_{4} e^{\Theta_1 + \Theta_1^* + \Theta_2 + \Theta_2^*} \\
	\label{s21}
	s_1 = \gamma_1 e^{\Theta_1} + \gamma_2 e^{\Theta_2}\\
	\label{s23}
	s_3 = \gamma_{12} e^{\Theta_1 + \Theta_1^* + \Theta_2} + \gamma_{21} e^{\Theta_1 + \Theta_2 + \Theta_2^*}
\end{align}
where $\Theta_j = p_j \xi + \frac{1}{p_j} \int a(\tau) d\tau$ for $j = 1, 2$. $p_j$'s are spectral parameters. Putting the above equations into Eqs. \ref{BR1} - \ref{BR3}, we get the following expressions
\begin{align}
	\label{betajk4}
	\beta_{jk} &= i\ c_{jk}^2\ p_j, \quad \quad \quad \quad \quad \quad \quad \beta_4 = -\frac{c_2^4\ p_1^*\ p_2^* |p_1 - p_2|^4}{(p_1 + p_1^*)^2 (p_2 + p_2^*)^2 |p_1 + p_2^*|^4} \\
	\label{alphajjk}
	\alpha_j &= \sqrt{2} c_{jj} (1 + \frac{p_j^*}{p_j}), \quad \quad \quad \alpha_{jk} = -\frac{2 i\ \beta_{jj}\ \beta_{kj}\ p_j^* (p_j - p_k)^2}{p_j^2\ p_k^2\ \alpha_j^*}
\end{align}
where $c_{jk} = \frac{c_2}{p_j + p_k}$ for $j, k = 1, 2$ and $c_2$ is an arbitrary real or purely imaginary constant. With this information we can realise the 2SS of DLFLE and consequently that of FLE. Just like for 1SS, here also we graphically represent 2SS of Eq. \ref{DLFLE} for three cases of the varying coefficient. 1. Figure \ref{Fig2a}: $a(\tau) = 1$ represents constant varying coefficient, i.e. no variation in dispersion and nonlinearity of the medium, here the two individual solitons move straight in their respective paths and only suffer a small shift in their phase upon interaction with each other and the rest of the characteristics of the solitons namely, size, shape, velocity, amplitude, etc. remains the same as before the interaction, this is one of the fundamental properties of the solitons that upon interaction with each other they suffer only a small shift in their phases and rest of the soliton characteristics remain the same, 2. Figure \ref{Fig2b}: $a(\tau) = cos(k \tau)$ represents sinusoidal nature of the dispersion and nonlinearity which results for the 2SS to have a sinusoidal propagation since their individual velocities are functions of $a(\tau)$, from the figure it is clearly seen that the individual solitons repeatedly suffer phase shift upon their repetitive interactions and 3. Figure \ref{Fig2c}: $a(\tau) = e^{\sigma \tau} cos(k \tau)$ is same as the previous Figure \ref{Fig2b} with $a(\tau)$ having an additional exponential function along with the cosine function, this makes the individual solitons deviate from the mean position more and more along the $\tau$ axis.

\vspace{2mm}
The width of the jth soliton is expressed as $\frac{2}{p_j + p_j^*}$ and the velocity is $\frac{a(\tau)}{|p_j|^2}$. Just like in the case of 1SS, the characteristics of te jth soliton of 2SS is similar. In 2SS also the choice of $a(\tau)$ gives us a control over the soliton management to alter the soliton dynamics just like in 1SS case.
\vspace{2mm}

From here, one may further derive the N soliton solution of DLFLE and FLE by considering the terms in $g$, $f$ and $s$ upto the order of $\epsilon^{2N+1}$ and putting them into Eqs. \ref{BR1} - \ref{BR3}.

\section{Conclusion}
\justifying
Hirota bilinearisation is very useful technique to derive soliton solutions of partial differential equations representing nonlinear systems and when introduced an auxiliary function, the bilinear process becomes much straight forward. The inclusion of the varying parameter makes the way to analyse the DLFLE or FLE system with varying dispersion and nonlinearity. The varying coefficient provides a great control over the propagation of the soliton and this information can be useful while designing an optical system for ultrashort pulse or plasma for ion-cyclotron wave to create desired wave. Our work can be further useful to study two component coupled DLFLE and FLE system and other nonlinear systems in various branches of physics having different varying coefficients for dispersion and nonlinearity. We believe our analysis will help the researchers to have further insights into DLFLE/FLE systems and our method may inspire researchers in obtaining nonlinear structures of various nonlinear medium.

\section*{Acknowledgment}
\justifying
R Dutta and S Talukdar acknowledge DST, Govt. of India for INSPIRE fellowship award. Corresponding award number DST/INSPIRE Fellowship/2020/IF200303 and DST/INSPIRE Fellowship/2020/IF200278.

\end{document}